\RequirePackage{lineno}
\documentclass[aps,twocolumn,showpacs,byrevtex,prl,reprint,nofootinbib]{revtex4-1}
\usepackage{xspace}

\usepackage{graphicx}% Include figure files
\usepackage{dcolumn}% Align table columns on decimal point
\usepackage{bm}% bold math
\usepackage{rotating}
\usepackage{color}
\usepackage{verbatim} %for comment
\usepackage{multirow}
\usepackage{amsmath}
\usepackage{relsize}
\usepackage{ulem}
\usepackage{mathrsfs}
\newcommand{\un}[1]{\ensuremath{\,\mathrm{#1}}}

\begin{document}
%\linenumbers

\title{\boldmath Evidence of a resonant structure in the $e^+e^-\to \pi^+D^0D^{*-}$ cross section between 4.05 and 4.60 GeV}
%2017.04.24
\author{
  \small
M.~Ablikim$^{1}$, M.~N.~Achasov$^{9,d}$, S. ~Ahmed$^{14}$, M.~Albrecht$^{4}$, A.~Amoroso$^{53A,53C}$, F.~F.~An$^{1}$, Q.~An$^{50,40}$, Y.~Bai$^{39}$, O.~Bakina$^{24}$, R.~Baldini Ferroli$^{20A}$
% ...
\\
\vspace{0.2cm}
(BESIII Collaboration)\\
\vspace{0.2cm} {\it
$^{1}$ Institute of High Energy Physics, Beijing 100049, People's Republic of China\\
% ...
$^{4}$ Bochum Ruhr-University, D-44780 Bochum, Germany\\
% ...
$^{9}$ G.I. Budker Institute of Nuclear Physics SB RAS (BINP), Novosibirsk 630090, Russia\\
% ...
$^{14}$ Helmholtz Institute Mainz, Johann-Joachim-Becher-Weg 45, D-55099 Mainz, Germany\\
% ...
$^{20}$ (A)INFN Laboratori Nazionali di Frascati, I-00044, Frascati, Italy; (B)INFN and University of Perugia, I-06100, Perugia, Italy\\
% ...
$^{24}$ Joint Institute for Nuclear Research, 141980 Dubna, Moscow region, Russia\\
% ...
$^{39}$ Southeast University, Nanjing 211100, People's Republic of China\\
$^{40}$ State Key Laboratory of Particle Detection and Electronics, Beijing 100049, Hefei 230026, People's Republic of China\\
% ...
$^{50}$ University of Science and Technology of China, Hefei 230026, People's Republic of China\\
% ...
$^{53}$ (A)University of Turin, I-10125, Turin, Italy; (B)University of Eastern Piedmont, I-15121, Alessandria, Italy; (C)INFN, I-10125, Turin, Italy\\
% ...
$^{d}$ Also at the Novosibirsk State University, Novosibirsk, 630090, Russia\\
% ...
}
}

\date{\today}

\begin{abstract}
  The cross section of the process $e^+e^-\to \pi^+D^0D^{*-}$ for center-of-mass energies from 4.05 to 4.60~GeV is measured precisely
  using data samples collected with the BESIII detector operating at the BEPCII storage ring.
  Two enhancements are clearly visible in the cross section around 4.23 and 4.40~GeV.
  Using several models to describe the dressed cross section yields stable parameters for the first enhancement,
  which has a mass of $4228.6 \pm 4.1 \pm 6.3 \un{MeV}/c^2$ and a width of $77.0 \pm 6.8 \pm 6.3 \un{MeV}$,
  where the first uncertainties are statistical and the second ones are systematic.
  Our resonant mass is consistent with previous observations of the $Y(4220)$ state
  and the theoretical prediction of a $D\bar{D}_1(2420)$ molecule.
  This result is the first observation of $Y(4220)$ associated with an open-charm final state.
  Fits with three resonance functions with additional $Y(4260)$,
  $Y(4320)$, $Y(4360)$, $\psi(4415)$, or a new resonance, do not
  show significant contributions from either of these resonances.
  The second enhancement is not from a single known resonance. It
  could contain contributions from $\psi(4415)$ and other resonances,
  and a detailed amplitude analysis is required to better understand this enhancement.
\end{abstract}


\maketitle

As the first observed charmonium-like state with $J^{\rm{PC}}=1^{--}$, the $Y(4260)$ has remained a mystery.
Many experimental measurements and theoretical interpretations have been proposed for this state~\cite{PhysRep},
such as hybrids~\cite{hybrid}, tetraquarks~\cite{tetraquark}, and hadronic molecules~\cite{molecule}.
Since it was observed only in hidden-charm processes, while its mass is close to open-charm thresholds,
studies of the open-charm production cross section in $e^+e^-$ annihilation will provide important information on its properties.
The cross section for $e^+e^-$ annihilation into $D^{(*)}\bar{D}^{(*)}$ pairs shows a dip at the resonance mass, 4.26~GeV/$c^2$~\cite{PRD80-072001}.
The $Y(4260)$ mass is only about 29~MeV/$c^2$ below the nominal threshold for $D\bar{D}_1(2420)$,
which is the first open-charm relative $S$-wave channel coupling to $J^{\rm{PC}} = 1^{--}$.
The $D\bar{D}_1(2420)$ molecule model is proposed as an interpretation of the $Y(4260)$,
but it predicts a significantly smaller mass of about 4.22~GeV/$c^2$~\cite{PRD90-074039, PRD94-054035}.

Recently, the precise measurement of the production cross section for $e^+e^-\to \pi^+\pi^-J/\psi$ from the BESIII experiment~\cite{PRL118-092001}
indicates that the structure around 4260~MeV/$c^2$ actually consists of two resonances with masses 4222~MeV/$c^2$ and 4320~MeV/$c^2$.
The mass of the former resonance (referred to as $Y(4220)$ hereafter) is consistent with the prediction of the $D\bar{D}_1(2420)$ molecule model.
Furthermore, a $Y(4220)$ resonance has also been reported by the BESIII collaboration in the cross-section measurements of $e^+e^-\to \omega\chi_{c0}$~\cite{PRL114-092003},
$e^+e^-\to \pi^+\pi^-h_c$~\cite{PRL118-092002}, and $e^+e^-\to \pi^+\pi^-\psi(3686)$~\cite{PRD96-032004}.
In addition, a new resonant structure with mass around 4.39~GeV/$c^2$, the $Y(4390)$,
has been reported by BESIII in the reactions $e^+e^-\to \pi^+\pi^-h_c$~\cite{PRL118-092002} and $e^+e^-\to \pi^+\pi^-\psi(3686)$~\cite{PRD96-032004}.
The mass of the $Y(4390)$ is about 45~MeV/$c^2$ and 70~MeV/$c^2$ higher than those of the $Y(4360)$~\cite{pdg}
and the second component of the $Y(4260)$ observed in $e^+e^-\to \pi^+\pi^-J/\psi$ by BESIII~\cite{PRL118-092001}, respectively.
The production of $e^+e^-\to \pi D\bar{D}^*$ is expected to be strongly enhanced above the nominal $D\bar{D}_1(2420)$ threshold
and could be a key for understanding existing puzzles with these $Y$ states~\cite{PRD94-054035}.

The cross section of $e^+e^- \to \pi^+D^0 D^{*-}$ was first measured by the Belle experiment using initial-state radiation (ISR)~\cite{PRD80-091101}.
No evidence for charmonium(-like) states was found within their statistics.
In this Letter, we report improved measurements of the production cross section of $e^+e^- \rightarrow \pi^+D^0D^{*-}$
at center-of-mass energies ($\sqrt{s}$) from 4.05 to 4.60~GeV using data samples taken
at 84 energy points~\cite{supplement} with the BESIII detector~\cite{bibbes3}.
The data set contains 5 energy points ($\sqrt{s} = 4.2263$, 4.2580, 4.3583, 4.4156, and 4.5995~GeV)
with integrated luminosities larger than 500~pb$^{-1}$ (\textquotedblleft H-XYZ data\textquotedblright \ hereafter),
and 79 energy points with integrated luminosities smaller than 200~pb$^{-1}$.
The $D^0$ meson is reconstructed in the $D^0 \rightarrow K^-\pi^+$ decay channel.
The bachelor $\pi^+$ produced directly in the $e^+e^-$ annihilation process is also reconstructed, while the $D^{*-}$ is not reconstructed directly,
but is inferred from energy-momentum conservation.
Charge-conjugate modes are implied, unless otherwise noted.

The BESIII detector is described in detail elsewhere~\cite{bibbes3}.
A Monte Carlo (MC) simulation based on {\footnotesize GEANT}4~\cite{bibgeant4} includes
the geometric description of the BESIII detector and its response.
For each energy point, we generate MC samples of the signal process, $e^+e^-\to \pi^+D^0D^{*-}$,
and the isospin partner process, $e^+e^-\to \pi^+D^-D^{*0}$, according to phase space (PHSP MC).
The effect of ISR is simulated with {\footnotesize KKMC}~\cite{kkmc} with a maximum energy for the ISR photon
corresponding to the $\pi^+D^0D^{*-}$ mass threshold.
Possible background contributions are estimated with {\footnotesize KKMC}-generated
`inclusive' MC samples with integrated luminosities comparable to the H-XYZ data,
where the known decay modes are simulated with {\footnotesize EVTGEN}~\cite{evtgen} using branching fractions taken from the PDG~\cite{pdg},
and the remaining unknown decays are simulated with the {\footnotesize LUNDCHARM} model~\cite{PRD62-034003}.

\begin{figure}[htbp]
  \centering
  \includegraphics[width=0.4\textwidth]{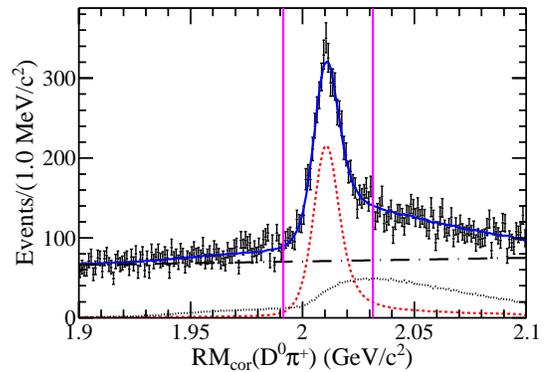}
  \caption{(Color online) Fit to the distribution of $RM_\text{cor}(D^0\pi^+)$ for the data sample at $\sqrt{s} = 4.5995$~GeV.
  The black dots with error bars are data, the solid line (blue) describes the total fit, the dashed line (red) describes
the signal shape, the dotted and dash-dotted lines (black) describe BKG1 and BKG2, respectively.
The pink vertical lines mark the signal region.}
  \label{fig::Dstmass}
\end{figure}

The charged tracks are reconstructed with standard selection requirements~\cite{PRD92-092006} and 
used to reconstruct $D^0$ meson candidates from $K^-\pi^+$ track pairs.
If there is more than one $D^0$ candidate in an event ($\sim$0.3\%), we choose the one whose invariant mass $M(K^-\pi^+)$ is closest to the world-average $D^0$ mass, $m(D^0)$~\cite{pdg}.
The signal region is defined as $|M(K^-\pi^+)-m(D^0)| < 15$~MeV/$c^2$.
To select the bachelor $\pi^+$, at least one extra charged track, which is not used in the $D^0$ candidate and has charge opposite to that of the reconstructed $K^-$,
is chosen with the same selection criteria as described above.
The $e^+e^-\to D^*\bar{D}^*$ background events are rejected by vetoing any $D^0\pi^+$ candidates satisfying $M(D^0\pi^+)<2.03$~GeV/$c^2$.
After the above requirements, the presence of a $D^{*-}$ meson is inferred by the invariant mass recoiling against the $D^0\pi^+$ system, $RM(D^0\pi^+)$.
To improve the mass resolution, the corrected recoil mass, $RM_\text{cor}(D^0\pi^+)=RM(D^0\pi^+)+M(K^-\pi^+)-m(D^0)$, is used, as shown in Fig.~\ref{fig::Dstmass}. If there is more than one bachelor $\pi^+$ in the event,
the one whose $RM_\text{cor}(D^0\pi^+)$ is closest to the world average $D^{*-}$ mass, $m(D^{*-})$, is selected.
A study of the inclusive MC samples shows that only the isospin partner process $e^+e^-\to \pi^+ D^- D^{*0}$ (BKG1, hereafter)
has an enhancement around the $D^{*-}$ mass region in the $RM_\text{cor}(D^0\pi^+)$ distribution.
The shape of this background process is different at each energy point and is taken from the MC simulation.
The $RM_\text{cor}(D^0\pi^+)$ distribution of the remaining background processes (BKG2, hereafter) does not peak and can be described by a first-order polynomial function.

An unbinned maximum likelihood fit to the $RM_\text{cor}(D^0\pi^+)$ distribution is performed to determine the signal yields.
The signal shape is derived from the MC shape convolved with a Gaussian function.
The background shape is parameterized as a sum of the shape from the PHSP MC sample for BKG1 and a first-order polynomial function for BKG2.
We perform a simultaneous fit to the $RM_\text{cor}(D^0\pi^+)$ distributions for all data samples to determine the yields of signal and background.
The mean values of the Gaussian smearing function are constrained to be the same for all energy points.
A center-of-mass energy-dependent width of the Gaussian function is obtained by fitting the widths
of the five H-XYZ data samples with a first-order polynomial function,
where these five widths are obtained by separate fits to the corresponding $RM_\text{cor}(D^0\pi^+)$ distributions.
The widths at $\sqrt{s}< 4.2263$~GeV are fixed to that at $\sqrt{s} = 4.2263$~GeV, since the fitted widths are close to zero.
Figure~\ref{fig::Dstmass} shows the fit result at $\sqrt{s} = 4.5995$~GeV.
The signal region is defined as $|RM_\text{cor}(D^0\pi^+) - \Delta M - m(D^{*-})|<20$~MeV/$c^2$,
where $\Delta M$ is the mean value of the Gaussian function obtained from the fit.
A sideband region, used below, is defined as $1.91<RM_\text{cor}(D^0\pi^+)<1.95$~GeV/$c^2$.

The Born cross sections ($\sigma_{\rm{Born}}$) and dressed cross sections ($\sigma_{\mathrm{dress}}$) at the individual energy points are calculated using
\begin{multline}
\sigma_{\rm{Born}}=\sigma_{\mathrm{dress}}\cdot|1-\Pi|^2
\\=\frac{N^{\rm{obs}}}{\mathcal{L}\cdot(1+\delta)\cdot\frac{1}{|1-\Pi|^2}\cdot\mathcal{B}(D^0\to K^-\pi^+)\cdot\epsilon},
\label{eq1}
\end{multline}
where $N^{\rm{obs}}$ is the signal yield, $\mathcal{L}$ is the integrated luminosity~\cite{CPC40-063001},
$1+\delta$ is the ISR correction factor~\cite{EAVS},
$\frac{1}{|1-\Pi|^2}$ is the correction factor for vacuum polarization~\cite{EPJC-66-585},
$\mathcal{B}(D^0\to K^-\pi^+) = (3.93 \pm 0.04)\%$~\cite{pdg}, and $\epsilon$ is the detection efficiency.
Values of all above variables are given in the supplemental material~\cite{supplement}.
Efficiencies at $\sqrt{s} = 4.2263$, 4.2580, 4.3583, 4.4156 and 4.5995~GeV are calculated with MC simulated data samples~\cite{PRD89-112006}
that are generated by the data-driven {\footnotesize BODY3} generator based on {\footnotesize EVTGEN}~\cite{evtgen},
taking into account the influence of possible intermediate states
($Z_c(3885)^-$ in the $D^0D^{*-}$ system~\cite{PRL112-022001, PRD92-092006} and highly-excited $D$ states in the $\pi^+D^0$ or $\pi^+D^{*-}$ systems).
Since the {\footnotesize BODY3} generator requires a large selected sample
obtained from events in the signal region after subtracting the background contribution (estimated with the events in the sideband region for BKG2 and MC simulation from BKG1),
it is only used for the five energy points with high luminosity.
Efficiencies at the other energy points are estimated with PHSP MC samples, with appropriate uncertainties included later.
The obtained Born cross sections, which are consistent with and more precise than those of Belle~\cite{PRD80-091101},
are summarized in the supplemental material~\cite{supplement}.

The systematic uncertainties in the cross section measurements are listed in Table~\ref{tab::sys}.
The uncertainty in luminosity is 1.0\% at each energy point~\cite{CPC39-093001}.
The uncertainty in $\mathcal{B}(D^0 \to K^- \pi^+)$ is 1.0\%~\cite{pdg}.
The uncertainty in the ISR correction factor is 3.0\%~\cite{EAVS}.
The uncertainties associated with the detection efficiencies include tracking
and PID efficiencies (1.0\% per track),
$D^0$ and $D^{*-}$ mass window requirements, and signal MC model.
The uncertainties associated with the $D^0$ and $D^{*-}$ mass windows are estimated by repeating the analysis with an altered mass window requirement;
the relative changes in the cross sections are taken as systematic uncertainties.
The uncertainties associated with the {\footnotesize BODY3} signal MC model
consist of three parts: the choice of binning, and the BKG1 and BKG2 subtractions.
The uncertainty associated with the choice of binning is estimated by repeating the simulation with an altered bin size.
The uncertainty associated with the BKG1 subtraction is studied by replacing the PHSP MC sample with MC samples
of processes including the intermediate states $e^+e^-\to \bar{D}_2^*(2460)^0D^{*0}$, $\bar{D}_2^*(2460)^0\to \pi^+D^-$ and
$e^+e^-\to D_1(2460)^+D^{-}$, $D_1(2460)^+\to \pi^+D^{*0}$.
For the BKG2 uncertainty, we replace the sideband events with the inclusive MC sample when subtracting the background.
The maximum relative changes on the detection efficiency are taken as the corresponding uncertainties.
The total signal MC model uncertainty is the sum in quadrature of these three contributions.
To estimate the uncertainties of the signal MC model for the low-luminosity data,
we estimate the detection efficiencies for the five energy points of large luminosity with the PHSP MC samples;
the resultant largest difference with respect to the nominal efficiencies, 5.3\%, is assigned as the corresponding uncertainty for the low-luminosity energy points.
The uncertainties associated with signal shape, background shape, and fit range in the signal yield extraction are determined by changing the signal shape to the pure MC shape,
by changing the background function from a linear polynomial function to a second-order one, and by changing the fit range, respectively.
Due to limited statistics, fit results at the energy points with low luminosity suffer large statistical fluctuations in such re-fits; thus,
the largest systematic uncertainties from the five large luminosity data samples are adopted.
Assuming no significant correlations between sources, the total systematic uncertainty is obtained as the sum in quadrature.

\begin{table}[htbp]
\begin{center}
\caption{
Breakdown of the systematic uncertainties (\%) in the measurements of the Born cross section,
separately for the five energy points with high luminosity data and the other points.
Part of the systematic uncertainties is in fact due to the finite statistics of the data.
}
\begin{tabular}{ccccccc}
\hline
\hline
$\sqrt{s}$ (GeV)  & 4.2263 & 4.2580 & 4.3583 & 4.4156 & 4.5995 & Other \\
\hline
Luminosity                       &     1.0 & 1.0 & 1.0 & 1.0 & 1.0 & 1.0 \\
$\mathcal{B}(D^0 \to K^- \pi^+)$ &     1.0 & 1.0 & 1.0 & 1.0 & 1.0 & 1.0 \\
$(1+\delta)\epsilon$    &     3.0 & 3.0 & 3.0 & 3.0 & 3.0 & 3.0 \\
Tracking                         &     3.0 & 3.0 & 3.0 & 3.0 & 3.0 & 3.0 \\
PID                              &     3.0 & 3.0 & 3.0 & 3.0 & 3.0 & 3.0 \\
$D^0$ mass window    		     &     0.3 & 0.1 & 0.4 & 0.2 & 0.7 & 0.7 \\
$D^{*-}$ mass window   		     &     0.2 & 0.1 & 0.2 & 0.3 & 0.3 & 0.3 \\
Signal MC model              &     2.5 & 2.1 & 2.9 & 2.3 & 2.2 & 5.3 \\
Signal shape                     &     0.1 & 1.5 & 0.8 & 1.5 & 2.1 & 2.1 \\
Background shape                 &     0.4 & 0.2 & 0.2 & 0.1 & 0.1 & 0.4 \\
Fit range                        &     0.1 & 0.2 & 0.1 & 0.1 & 0.1 & 0.2 \\
\hline
Sum in quadrature                &     6.0 & 6.0 & 6.2 & 6.1 & 6.2 & 8.0 \\
\hline\hline
\end{tabular}
\label{tab::sys}
\end{center}
\end{table}

\begin{figure}[htbp]
\includegraphics[width=0.48\textwidth]{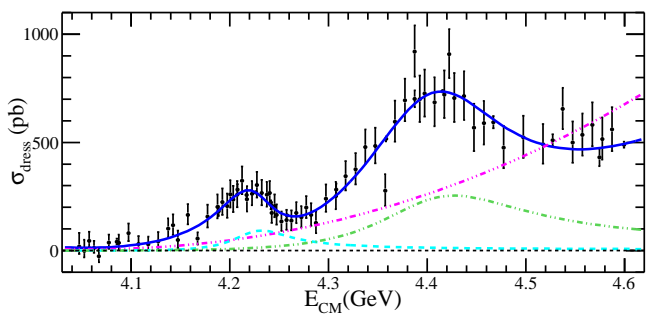}
\caption{(Color online) Fit to the dressed cross section of $e^+e^-\to \pi^+D^0D^{*-}$,
where the black dots with error bars are the measured cross sections and the blue line shows the fit result.
The error bars are statistical only.
The pink dashed triple-dot line describes the phase-space contribution, the green dashed double-dot line describes the $R_2$ contribution,
and the light blue dashed line describes the $R_1$ contribution.}
\label{fig::cros}
\end{figure}

The dressed cross section, which includes vacuum polarization effects, is shown in Fig.~\ref{fig::cros}.
Two enhancements around 4.23 and 4.40~GeV, denoted hereafter as $R_1$ and $R_2$, are clearly visible.
A maximum likelihood fit to the dressed cross section is performed to determine the parameters of these two enhancements.
Since the measured cross sections have asymmetric uncertainties for the data with low statistics,
the likelihood is described by an asymmetric Gaussian function as discussed in Ref.~\cite{PRL118-092002}.
In the fit, the total amplitude is described by the coherent sum of
a direct three-body phase-space term for $e^+e^-\to \pi^+D^0D^{*-}$ and
two relativistic Breit-Wigner (BW) functions,
representing the resonant structures $R_1$ and $R_2$:
\begin{widetext}
\begin{equation}
\label{crospdf}
\sigma_{\mathrm{dress}}(m) = \left|c\cdot \sqrt{P(m)}+e^{i\phi_1}B_1(m)\sqrt{P(m)/P(M_1)}+e^{i\phi_2}B_2(m)\sqrt{P(m)/P(M_2)}\right|^2,
\end{equation}
\end{widetext}
where the three-body phase-space factor $P(m)$~\cite{pdg} is modeled as a fixed fourth-order polynomial function.
The factor $B_{j}(m)=\frac{\sqrt{12\pi\Gamma^{\rm{el}}_{j}\Gamma_{j}}}{m^2-M_{j}^2+iM_{j}\Gamma_{j}}$
with $j = 1$ or 2 is the BW function for a vector state,
where $\Gamma^{\rm{el}}_{j} = \Gamma_{\rm{e^+e^-}}\mathcal{B}(\pi^+D^0D^{*-})_{j}$
is the product of the electronic partial width and the branching fraction to $\pi^+D^0D^{*-}$.
Parameters $c$, $M_{j}$, $\Gamma_{j}$, $\Gamma^{\rm{el}}_{j}$, and $\phi_{j}$ are the free parameters in the nominal fit.
The beam energy spread (1.6~MeV) is considered by convolving with a Gaussian function whose width is 1.6~MeV.
Only statistical uncertainties on the dressed cross sections are considered in the fit.
There are four solutions with the same fit quality~\cite{ZhuK} and identical resonance parameters for $R_1$ and $R_2$,
but different $c$, $\Gamma^{\rm{el}}_{j}$ and $\phi_{j}$, as listed in Table~\ref{tab:res}.
We also fit the dressed cross section with the coherent sum of one resonance and a phase-space term;
the change of the likelihood value, $\Delta(-2\rm{ln}L)$, with respect to that of nominal fit including two resonances is 124.3.
Taking into account the change in the number of degrees of freedom (4),
the statistical significance of the two-resonance model
over the one-resonance model is estimated to be 10.5$\sigma$.

Belle has observed the $\psi(4415)$ in $\psi(4415) \rightarrow D\bar{D}_2^*(2460)$~\cite{PRL100-062001} which can also decay to the
final state considered in this analysis.
Considering the observations of other charmonium(-like) states, models fixing the mass and width of $R_2$ to those of $Y(4260)$, $Y(4320)$, $Y(4360)$, or $\psi(4415)$
are also investigated and ruled out with a confidence level equivalent to more than 5.0$\sigma$.
Models including one additional known resonance, either $Y(4260)$, $Y(4320)$, $Y(4360)$, or $\psi(4415)$,
in which the masses and widths of these resonances are fixed to the world average values~\cite{pdg}, can improve the fit quality.
However, the statistical significances of the additional resonances are only 0.4$\sigma$, 0.4$\sigma$, 1.4$\sigma$, and 1.0$\sigma$, respectively.
The statistical significance of an additional unknown resonance is only 0.8$\sigma$,
accounting for the two extra free parameters of mass and width.
The high-mass enhancement has a more complicated underlying
structure, the understanding of which requires a detailed amplitude
analysis that is beyond the scope of this Letter.

All above models yield a stable set of parameters for $R_1$ but wildly varying parameters for $R_2$,
so we only estimate the systematic uncertainties of parameters of $R_1$,
which are mainly from the uncertainties of the absolute center-of-mass energy measurements,
the cross section measurements, and the parameterization of the three-body phase-space factor.
The uncertainty of the energy measurement (0.8~MeV) is propagated to the masses of the resonances.
The uncertainty associated with the cross section measurements consists of two parts.
The first is from the common uncertainties of the measured cross sections (tracking, PID, luminosity,
and $\mathcal{B}(D^0 \to K^- \pi^+)$) for all energy points (4.5\%);
we shift up or down all measured cross sections by 4.5\% simultaneously and repeat the fit on the measured cross sections.
The differences, 0.1~MeV/$c^2$ for the mass and 0.1~MeV for the width,
are taken as systematic uncertainties for the $R_1$ resonance.
The second part includes all the other uncertainties of the measured cross sections.
We add these uncertainties into the statistical ones in quadrature, and repeat the fit.
The resulting differences in resonance parameters, 4.9~MeV/$c^2$ for the mass and 2.7~MeV for the width of $R_1$, are taken as systematic uncertainties.
The uncertainty associated with the three-body phase-space factor is determined by changing the parameterization function
from a fourth-order polynomial function to a third-order one.
The resulting differences, 3.8~MeV/$c^2$ for the mass and 5.7~MeV for the width, are taken as the corresponding uncertainties for $R_1$.
Assuming the individual systematic uncertainties are uncorrelated, the total systematic uncertainty is obtained by summing the individual values in quadrature,
yielding 6.3~MeV/$c^2$ for the mass and 6.3~MeV for the width of $R_1$.

\begin{table}[htbp]
\begin{center}
\caption{The fitted parameters of the cross sections of $e^+e^-\to \pi^+D^0D^{*-}$.
The uncertainties are statistical only.}
\smaller
\begin{tabular}{ccccc}
\hline
\hline
Parameter  & Solution I & Solution II & Solution III & Solution IV \\
\hline
$c$ (MeV$^{-3/2}$)     & \multicolumn{4}{c}{$(6.2 \pm 0.5)\times10^{-4}$}    \\
$M_1$ (MeV/$c^2$)   & \multicolumn{4}{c}{$4228.6 \pm 4.1$} \\
$\Gamma_{1}$ (MeV)  & \multicolumn{4}{c}{$77.0 \pm 6.8$}   \\
$M_2$ (MeV/$c^2$)   & \multicolumn{4}{c}{$4404.7 \pm 7.4$} \\
$\Gamma_{\rm{2}}$ (MeV)  & \multicolumn{4}{c}{$191.9 \pm 13.0$} \\
$\Gamma^{\rm{el}}_1$ (eV) & $77.4 \pm 10.1$   & $8.6 \pm 1.6$   & $99.5 \pm 14.6$   & $11.1 \pm 2.3$ \\
$\Gamma^{\rm{el}}_2$ (eV) & $100.4 \pm 13.3$  & $64.2 \pm 8.0$  & $664.2 \pm 80.0$  & $423.0 \pm 47.0$ \\
$\phi_1$  (rad)           & $-2.0 \pm 0.1$   & $3.0 \pm 0.2$   & $-0.9 \pm 0.1$    & $-2.2 \pm 0.1$ \\
$\phi_2$  (rad)           & $2.1 \pm 0.2$    & $2.5 \pm 0.2$   & $-2.3 \pm 0.1$    & $-1.9 \pm 0.1$ \\
\hline
\hline
\end{tabular}
\label{tab:res}
\end{center}
\end{table}

In summary, the Born cross section for the process $e^+e^- \rightarrow \pi^+D^0D^{*-}$ is precisely measured
using the data samples collected at 84 energy points from 4.05 to 4.60~GeV with the BESIII detector.
Two enhancements are observed in the dressed cross sections around 4.23 and 4.40~GeV.
Using many models to describe the dressed cross section,
we obtain a stable resonant structure around 4.23~GeV, the parameters
of which are measured to be $M(R_1) = (4228.6 \pm 4.1 \pm 6.3)$~MeV/$c^2$ and $\Gamma(R_1) = (77.0 \pm 6.8 \pm 6.3)$~MeV,
where the first uncertainties are statistical and the second ones systematic.
The resonance parameters for the enhancement around 4.40~GeV are strongly dependent on the model assumptions,
necessitating further studies.

The statistical significance of the two-resonance model over a one-resonance model is estimated to be 10.5$\sigma$.
This is the first experimental evidence for open-charm production associated with the $Y$ states.
The mass of $R_1$ is consistent with the mass of the resonance observed in the hidden-charm processes by the BESIII experiment
as well as the theoretical prediction of the $D\bar{D}_1(2420)$ molecule interpretation~\cite{PRD90-074039}.
The width of $R_1$ is consistent with that of $e^+e^- \rightarrow \pi^{+}\pi^{-}h_c$~\cite{PRL118-092002} and
$e^+e^-\to \pi^+\pi^-\psi(3686)$~\cite{PRD96-032004},
but it is about 39 and 33~MeV/$c^2$ higher than that seen in $e^+e^-\to \omega\chi_{c0}$~\cite{PRL114-092003} and
$e^+e^-\to \pi^+\pi^-J/\psi$~\cite{PRL118-092001}, respectively.
The minimum and maximum of the branching ratio, $\frac{\mathcal{B}(Y(4220)\to\pi D\bar{D}^{*})}{\mathcal{B}(Y(4220)\to\pi\pi J/\psi)}$
($\frac{\mathcal{B}(Y(4220)\to\pi D\bar{D}^{*})}{\mathcal{B}(Y(4220)\to\pi\pi h_c)}$),
are calculated to be $1.3 \pm 0.3$ and $124.3 \pm 36.1$ ($3.7^{+2.5}_{-1.5}$ and $43.3^{+29.0}_{-16.4}$) by assuming isospin symmetry, respectively.
The measured Born cross section of $e^+e^- \rightarrow \pi^+D^0D^{*-}$ at the $Y(4220)$ peak is higher than the sum of the
known hidden-charm channels. Since no other open-charm production associated with this $Y$ state has yet been reported,
the $\pi^+D^0D^{*-}$ final state may be the dominant decay mode of the $Y(4220)$ state,
as predicted by the $D\bar{D}_1(2420)$ molecule interpretation~\cite{PRD90-074039}.
No significant contributions from a third resonance are observed using three-resonance models with additional
$Y(4260)$, $Y(4320)$, $Y(4360)$, $\psi(4415)$, or a new resonance,
while $Y(4320)$ and $\psi(4415)$ are observed in $e^+e^- \rightarrow \pi^{+}\pi^{-}J/\psi$~\cite{PRL118-092001} and
$\psi(4415) \rightarrow D\bar{D}_2^*(2460)$~\cite{PRL100-062001}, respectively.
The amplitude studies of this final state and more studies on other open-charm production modes
will shed additional light on the nature of these charmonium(-like) states.

\par

The BESIII collaboration thanks the staff of BEPCII and the IHEP computing center for their strong support.
This work is supported in part by National Key Basic Research Program of China under Contract No. 2015CB856700;
National Natural Science Foundation of China (NSFC) under Contracts Nos. 11521505, 11075174, 11121092, 11235011, 11335008, 11405046, 11425524, 11475185, 11575077, 11605042, 11625523, 11635010;
Chinese Academy of Science Focused Science Grant;
the Chinese Academy of Sciences (CAS) Large-Scale Scientific Facility Program; the CAS Center for Excellence in Particle Physics (CCEPP);
Joint Large-Scale Scientific Facility Funds of the NSFC and CAS under Contracts Nos. U1332201, U1532257, U1504112, U1532258, U1632109;
CAS under Contracts Nos. KJCX2-YW-N29, KJCX2-YW-N45;
CAS Key Research Program of Frontier Sciences under Contract No. QYZDJ-SSW-SLH003, QYZDJ-SSW-SLH040;
100 Talents Program of CAS;
National 1000 Talents Program of China;
INPAC and Shanghai Key Laboratory for Particle Physics and Cosmology;
German Research Foundation DFG under Contracts Nos. Collaborative Research Center CRC 1044, FOR 2359;
Istituto Nazionale di Fisica Nucleare, Italy;
Koninklijke Nederlandse Akademie van Wetenschappen (KNAW) under Contract No. 530-4CDP03;
Ministry of Development of Turkey under Contract No. DPT2006K-120470;
National Science and Technology fund; The Swedish Research Council;
U. S. Department of Energy under Contracts Nos. DE-FG02-05ER41374, DE-SC-0010118, DE-SC-0010504, DE-SC-0012069;
University of Groningen (RuG) and the Helmholtzzentrum fuer Schwerionenforschung GmbH (GSI), Darmstadt;
WCU Program of National Research Foundation of Korea under Contract No. R32-2008-000-10155-0;
China Postdoctoral Science Foundation.

\par

\clearpage
\newpage
\begin{appendix}

\begin{widetext}
\section{Supplemental material}

The integrated luminosities $\mathcal{L}$, the numbers of signal events $N^{\rm{obs}}$, the ISR correction factors $1+\delta$,
the correction factors from vacuum polarization $\frac{1}{|1-\Pi|^2}$,
the detection efficiencies $\epsilon$, and the Born cross sections $\sigma_{\rm{Born}}$ are summarized in Table~\ref{tab::sup1} and Table~\ref{tab::sup2}.
15 energy points with integrated luminosities larger than 40~pb$^{-1}$ are called \textquotedblleft $XYZ$ data\textquotedblright \ ,
and 69 energy points with integrated luminosities smaller than 20~pb$^{-1}$ called \textquotedblleft $R$-scan data\textquotedblright \ .

\begin{table}[htbp]
\begin{center}
\caption{The Born cross sections of $e^+e^-\to \pi^+D^0D^{*-}$ for $XYZ$ data. The first uncertainties are statistical and the second ones systematic.}
\begin{tabular}{ccccccc}
\hline
\hline
$\sqrt{s}$ (GeV)  &  $\mathcal{L}$ (pb$^{-1}$)  &  $N^{\rm{obs}}$  &  $1+\delta$  & $\frac{1}{|1-\Pi|^2}$  &  $\epsilon$ (\%)  &  $\sigma_{\rm{Born}}$ (pb) \\
\hline
4.0855      & 52.6           &  19$\pm$6                     & 0.725  & 1.06  & 30.05  &  40$\pm$13$\pm$3    \\
4.1886      & 43.1           &  95$\pm$12                    & 0.749  & 1.07  & 41.05  &  171$\pm$22$\pm$14  \\
4.2077      & 54.6           &  190$\pm$16                   & 0.754  & 1.07  & 41.73  &  263$\pm$22$\pm$21  \\
4.2171      & 54.1           &  175$^{+16}_{-15}$            & 0.765  & 1.07  & 41.86  &  241$^{+22}_{-21}\pm$19  \\
4.2263      & 1091.7         &  3885$\pm$72                  & 0.786  & 1.07  & 42.82  &  252$\pm$5$\pm$15    \\
4.2417      & 55.6           &  157$^{+16}_{-15}$            & 0.858  & 1.06  & 39.64  &  199$^{+20}_{-19}\pm$16  \\
4.2580      & 825.7          &  1817$\pm$56                  & 0.903  & 1.06  & 38.17  &  153$\pm$5$\pm$9     \\
4.3079      & 44.9           &  162$\pm$16                   & 0.813  & 1.06  & 40.11  &  267$\pm$26$\pm$21  \\
4.3583      & 539.8          &  3788$\pm$80                  & 0.787  & 1.05  & 43.95  &  491$\pm$10$\pm$30  \\
4.3874      & 55.2           &  510$\pm$28                   & 0.798  & 1.05  & 42.01  &  666$\pm$37$\pm$53  \\
4.4156      & 1073.6         &  10899$\pm$142                & 0.821  & 1.05  & 42.76  &  698$\pm$9$\pm$43          \\
4.4671      & 109.9          &  871$^{+42}_{-41}$            & 0.887  & 1.06  & 38.34  &  559$^{+27}_{-26}\pm$45  \\
4.5271      & 110.0          &  748$^{+41}_{-40}$            & 0.931  & 1.06  & 36.46  &  481$^{+26}_{-26}\pm$38  \\
4.5745      & 47.7           &  271$^{+26}_{-25}$            & 0.925  & 1.06  & 36.37  &  406$^{+39}_{-37}\pm$32  \\
4.5995      & 566.9          &  3605$\pm$101                 & 0.916  & 1.06  & 36.06  &  462$\pm$13$\pm$29         \\
\hline
\hline
\end{tabular}
\label{tab::sup1}
\end{center}
\end{table}

\begin{table}[htbp]
\begin{center}
\caption{The Born cross sections of $e^+e^-\to \pi^+D^0D^{*-}$ for $R$-scan data. The first uncertainties are statistical and the second ones systematic.}
\begin{tabular}{ccccccc}
\hline
\hline
$\sqrt{s}$ (GeV)  &  $\mathcal{L}$ (pb$^{-1}$)  &  $N^{\rm{obs}}$  &  $1+\delta$  & $\frac{1}{|1-\Pi|^2}$  &  $\epsilon$ (\%)  &  $\sigma_{\rm{Born}}$ (pb) \\
\hline
4.0474   & 6.567   & 0.3$^{+1.0}_{-0.5}$     & 0.707  & 1.05   &  8.58  & 18$^{+60}_{-33}\pm$1 \\ 
4.0524   & 6.927   & $-$0.02$^{+1.2}_{-0.7}$  & 0.713  & 1.05   & 12.49 & $-$1.0$^{+48.3}_{-28.6}\pm$0.1    \\
4.0574   & 6.338   & 1.3$^{+1.2}_{-0.8}$     & 0.712  & 1.05   & 16.01  & 42$^{+40}_{-26}\pm$3     \\ 
4.0624   & 7.022   & 0.5$^{+1.2}_{-0.9}$     & 0.712  & 1.06   & 19.43  & 12$^{+31}_{-21}\pm$1     \\ 
4.0674   & 7.271   & $-$1.2$^{+1.6}_{-1.2}$  & 0.720  & 1.06   & 21.63  & $-$25$^{+35}_{-26}\pm$2   \\ 
4.0774   & 7.721   & 2.2$^{+2.1}_{-1.7}$     & 0.723  & 1.06   & 26.78  & 36$^{+34}_{-28}\pm$3     \\ 
4.0874   & 7.611   & 2.3$^{+2.5}_{-2.1}$     & 0.728  & 1.06   & 30.05  & 33$^{+37}_{-31}\pm$3     \\ 
4.0974   & 7.254   & 5.2$^{+2.8}_{-2.4}$     & 0.663  & 1.06   & 33.77  & 77$^{+42}_{-36}\pm$6     \\ 
4.1074   & 7.146   & 2.1$^{+2.6}_{-2.1}$     & 0.683  & 1.05   & 35.60  & 29$^{+36}_{-30}\pm$2     \\ 
4.1174   & 7.648   & 2.2$^{+2.8}_{-2.3}$     & 0.691  & 1.06   & 36.90  & 28$^{+34}_{-29}\pm$2     \\ 
4.1274   & 7.207   & 3.1$^{+3.3}_{-2.8}$     & 0.699  & 1.06   & 37.54  & 40$^{+42}_{-36}\pm$3     \\ 
4.1374   & 7.268   & 7.9$^{+3.6}_{-3.2}$     & 0.708  & 1.06   & 38.46  & 96$^{+44}_{-39}\pm$8     \\ 
4.1424   & 7.774   & 10.0$^{+4.2}_{-3.8}$    & 0.710  & 1.06   & 39.12  & 111$^{+47}_{-42}\pm$9    \\ 
4.1474   & 7.662   & 4.2$^{+3.6}_{-3.2}$     & 0.719  & 1.06   & 39.26  & 47$^{+40}_{-35}\pm$4     \\ 
4.1574   & 7.954   & 14.7$^{+4.4}_{-4.0}$    & 0.726  & 1.06   & 39.05  & 156$^{+47}_{-42}\pm$12   \\ 
4.1674   & 18.008  & 11.6$^{+6.3}_{-5.8}$    & 0.734  & 1.06   & 40.41  & 52$^{+28}_{-26}\pm$4     \\ 
4.1774   & 7.309   & 13.4$^{+4.7}_{-4.3}$    & 0.743  & 1.06   & 39.93  & 148$^{+52}_{-47}\pm$12   \\ 
4.1874   & 7.560   & 18.8$^{+5.3}_{-4.9}$    & 0.754  & 1.07   & 41.50  & 190$^{+54}_{-49}\pm$15   \\ 
4.1924   & 7.503   & 20.4$^{+5.8}_{-5.3}$    & 0.760  & 1.07   & 40.53  & 210$^{+60}_{-55}\pm$17   \\ 
4.1974   & 7.582   & 19.2$^{+5.4}_{-5.0}$    & 0.762  & 1.07   & 41.11  & 192$^{+54}_{-50}\pm$15   \\ 
4.2004   & 6.815   & 21.9$^{+5.5}_{-5.1}$    & 0.761  & 1.07   & 41.40  & 243$^{+61}_{-56}\pm$19   \\ 
4.2034   & 7.638   & 22.5$^{+6.0}_{-5.5}$    & 0.762  & 1.07   & 41.60  & 222$^{+59}_{-54}\pm$18   \\ 
\hline
\multicolumn{7}{r}{Continued on next page} \\
\hline
\end{tabular}
\label{tab::sup2}
\end{center}
\end{table}

\begin{table}[htbp]
\begin{center}
\caption{Table~\ref{tab::sup2}-continued from previous page.}
\begin{tabular}{ccccccc}
\hline
\hline
$\sqrt{s}$ (GeV)  &  $\mathcal{L}$ (pb$^{-1}$)  &  $N^{\rm{obs}}$  &  $1+\delta$  & $\frac{1}{|1-\Pi|^2}$  &  $\epsilon$ (\%)  &  $\sigma_{\rm{Born}}$ (pb) \\
\hline
4.2074     & 7.678   & 24.4$^{+5.9}_{-5.5}$    & 0.762  & 1.07   &  40.99   &  242$^{+59}_{-55}\pm$19   \\
4.2124     & 7.768   & 31.2$^{+6.4}_{-6.0}$    & 0.763  & 1.07   &  41.52   &  302$^{+62}_{-58}\pm$24   \\
4.2174     & 7.935   & 23.2$^{+5.9}_{-5.5}$    & 0.762  & 1.07   &  41.18   &  222$^{+56}_{-52}\pm$18   \\
4.2224     & 8.212   & 26.8$^{+5.9}_{-5.5}$    & 0.769  & 1.07   &  41.12   &  246$^{+54}_{-51}\pm$20   \\
4.2274     & 8.193   & 31.1$^{+6.3}_{-5.9}$    & 0.784  & 1.07   &  40.61   &  284$^{+58}_{-54}\pm$23   \\
4.2324     & 8.273   & 27.8$^{+6.3}_{-5.9}$    & 0.801  & 1.06   &  40.08   &  250$^{+56}_{-53}\pm$20   \\
4.2374     & 7.830   & 26.1$^{+6.0}_{-5.5}$    & 0.831  & 1.06   &  39.45   &  243$^{+55}_{-52}\pm$19   \\
4.2404     & 8.571   & 30.2$^{+6.3}_{-5.9}$    & 0.854  & 1.06   &  39.53   &  250$^{+52}_{-48}\pm$20   \\
4.2424     & 8.487   & 19.6$^{+5.6}_{-5.2}$    & 0.866  & 1.06   &  38.92   &  164$^{+47}_{-43}\pm$13   \\
4.2454     & 8.554   & 23.6$^{+6.3}_{-5.8}$    & 0.883  & 1.06   &  38.69   &  193$^{+51}_{-48}\pm$15   \\
4.2474     & 8.596   & 19.3$^{+5.7}_{-5.3}$    & 0.895  & 1.06   &  38.56   &  156$^{+46}_{-43}\pm$12   \\
4.2524     & 8.657   & 16.2$^{+5.6}_{-5.1}$    & 0.920  & 1.06   &  38.49   &  127$^{+44}_{-40}\pm$10   \\
4.2574     & 8.880   & 17.5$^{+5.9}_{-5.5}$    & 0.928  & 1.06   &  38.29   &  133$^{+45}_{-42}\pm$11   \\
4.2624     & 8.629   & 16.3$^{+5.7}_{-5.2}$    & 0.932  & 1.06   &  37.40   &  130$^{+45}_{-42}\pm$10   \\
4.2674     & 8.548   & 20.1$^{+5.9}_{-5.4}$    & 0.924  & 1.06   &  37.35   &  164$^{+48}_{-44}\pm$13   \\
4.2724     & 8.567   & 17.9$^{+6.0}_{-5.5}$    & 0.914  & 1.06   &  37.64   &  146$^{+49}_{-45}\pm$12   \\
4.2774     & 8.723   & 23.4$^{+6.1}_{-5.7}$    & 0.895  & 1.06   &  38.27   &  189$^{+50}_{-46}\pm$15   \\
4.2824     & 8.596   & 18.8$^{+6.2}_{-5.7}$    & 0.878  & 1.06   &  38.45   &  156$^{+51}_{-47}\pm$12   \\
4.2874     & 9.010   & 15.3$^{+6.2}_{-5.6}$    & 0.861  & 1.06   &  39.00   &  122$^{+49}_{-45}\pm$10   \\
4.2974     & 8.453   & 25.9$^{+7.0}_{-6.6}$    & 0.828  & 1.06   &  39.18   &  228$^{+62}_{-58}\pm$18   \\
4.3074     & 8.599   & 27.3$^{+7.0}_{-6.5}$    & 0.803  & 1.06   &  39.82   &  239$^{+62}_{-57}\pm$19   \\
4.3174     & 9.342   & 40.8$^{+8.3}_{-7.8}$    & 0.790  & 1.05   &  40.82   &  327$^{+66}_{-62}\pm$26   \\
4.3274     & 8.657   & 40.8$^{+8.1}_{-7.6}$    & 0.781  & 1.05   &  40.87   &  358$^{+71}_{-67}\pm$29   \\
4.3374     & 8.700   & 53.2$^{+9.0}_{-8.6}$    & 0.777  & 1.05   &  41.80   &  456$^{+77}_{-73}\pm$36   \\
4.3474     & 8.542   & 51.4$^{+8.9}_{-8.5}$    & 0.774  & 1.05   &  40.91   &  460$^{+80}_{-76}\pm$37   \\
4.3574     & 8.063   & 28.5$^{+7.8}_{-7.3}$    & 0.775  & 1.05   &  41.84   &  264$^{+72}_{-67}\pm$21   \\
4.3674     & 8.498   & 64.6$^{+10.5}_{-10.0}$  & 0.772  & 1.05   &  42.02   &  567$^{+92}_{-88}\pm$45   \\
4.3774     & 8.158   & 71.2$^{+10.3}_{-9.9}$   & 0.776  & 1.05   &  41.24   &  659$^{+96}_{-92}\pm$53   \\
4.3874     & 7.460   & 86.3$^{+11.4}_{-10.9}$  & 0.786  & 1.05   &  40.70   &  874$^{+115}_{-111}\pm$70 \\
4.3924     & 7.430   & 67.0$^{+10.5}_{-10.1}$  & 0.787  & 1.05   &  41.68   &  665$^{+104}_{-100}\pm$53 \\
4.3974     & 7.178   & 67.8$^{+10.4}_{-9.9}$   & 0.795  & 1.05   &  41.68   &  689$^{+106}_{-101}\pm$55 \\
4.4074     & 6.352   & 58.0$^{+9.8}_{-9.4}$    & 0.804  & 1.05   &  42.14   &  650$^{+110}_{-105}\pm$52 \\
4.4174     & 7.519   & 70.2$^{+10.9}_{-10.5}$  & 0.816  & 1.05   &  40.38   &  683$^{+106}_{-102}\pm$55 \\
4.4224     & 7.436   & 89.1$^{+11.4}_{-10.9}$  & 0.824  & 1.06   &  40.72   &  860$^{+110}_{-106}\pm$69 \\
4.4274     & 6.788   & 63.0$^{+10.3}_{-9.9}$   & 0.829  & 1.06   &  40.39   &  668$^{+110}_{-105}\pm$53 \\
4.4374     & 7.634   & 72.6$^{+11.7}_{-11.2}$  & 0.842  & 1.06   &  40.23   &  674$^{+109}_{-104}\pm$54 \\
4.4474     & 7.677   & 57.5$^{+11.3}_{-10.8}$  & 0.855  & 1.06   &  39.29   &  536$^{+105}_{-101}\pm$43 \\
4.4574     & 8.724   & 68.5$^{+12.0}_{-11.5}$  & 0.870  & 1.06   &  38.98   &  555$^{+97}_{-93}\pm$44   \\
4.4774     & 8.167   & 52.1$^{+11.0}_{-10.4}$  & 0.901  & 1.06   &  37.84   &  449$^{+94}_{-90}\pm$36   \\
4.4974     & 7.997   & 56.3$^{+10.7}_{-10.2}$  & 0.919  & 1.06   &  37.26   &  494$^{+94}_{-90}\pm$40   \\
4.5174     & 8.674   & 57.0$^{+10.8}_{-10.4}$  & 0.932  & 1.06   &  36.46   &  464$^{+88}_{-84}\pm$37   \\
4.5374     & 9.335   & 82.4$^{+12.2}_{-11.8}$  & 0.936  & 1.06   &  36.62   &  619$^{+92}_{-89}\pm$49   \\
4.5474     & 8.765   & 58.0$^{+11.3}_{-10.8}$  & 0.937  & 1.06   &  35.94   &  472$^{+92}_{-88}\pm$38   \\
4.5574     & 8.259   & 59.0$^{+10.9}_{-10.5}$  & 0.934  & 1.06   &  36.40   &  505$^{+94}_{-90}\pm$40   \\
4.5674     & 8.390   & 65.5$^{+11.6}_{-11.1}$  & 0.932  & 1.06   &  36.62   &  549$^{+97}_{-93}\pm$44   \\
4.5774     & 8.545   & 57.2$^{+11.0}_{-10.5}$  & 0.930  & 1.06   &  35.49   &  487$^{+94}_{-90}\pm$39   \\
4.5874     & 8.162   & 59.6$^{+11.0}_{-10.6}$  & 0.926  & 1.06   &  35.84   &  528$^{+97}_{-94}\pm$42   \\
\hline
\end{tabular}
\end{center}
\end{table}
\clearpage
\end{widetext}

\end{appendix}


\begin{thebibliography}{99}
\bibitem{PhysRep} H.~X.~Chen, W.~Chen, X.~Liu and S.~L.~Zhu, Phys. Rep. {\bf 639}, 1 (2016).
\bibitem{hybrid} S.~L.~Zhu, Phys. Lett. B {\bf 625}, 212 (2005); F.~E.~Close and P.~R.~Page, Phys. Lett. B {\bf 628}, 215 (2005);
E.~Kou and O.~Pene, Phys. Lett. B {\bf 631}, 164 (2005).
\bibitem{tetraquark} L.~Maiani, V.~Riquer, F. Piccinini, and A.~D.~Polosa, Phys. Rev. D {\bf 72}, 031502(R) (2005).
\bibitem{molecule} X.~Liu, X.~Q.~Zeng, and X.~Q.~Li, Phys. Rev. D {\bf 72}, 054023(R) (2005);
F.~K.~Guo, C.~Hanhart, U.~G.~Mei\ss{}ner, Q.~Wang, Q.~Zhao and B.~S.~Zou, Rev.\ Mod.\ Phys.\  {\bf 90}, 015004 (2018).
\bibitem{PRD80-072001} D.~Cronin-Hennessy {\it et al.} (CLEO Collaboration), Phys. Rev. D {\bf 80}, 072001 (2009).
\bibitem{PRD90-074039} M.~Cleven {\it et al}., Phys. Rev. D {\bf 90}, 074039 (2014).
\bibitem{PRD94-054035} Q.~Wang, C.~Hanhart and Q.~Zhao, Phys.\ Rev.\ Lett.\  {\bf 111}, 132003 (2013);
W.~Qin, S.~R.~Xue, and Q.~Zhao, Phys. Rev. D {\bf 94}, 054035 (2016).
\bibitem{PRL118-092001} M.~Ablikim {\it et al}.~(BESIII Collaboration), Phys. Rev. Lett. {\bf 118}, 092001 (2017).
\bibitem{PRL114-092003} M.~Ablikim {\it et al.} (BESIII Collaboration), Phys.\ Rev.\ Lett.\ {\bf 114}, 092003 (2015).
\bibitem{PRL118-092002} M.~Ablikim {\it et al}.~(BESIII Collaboration), Phys. Rev. Lett. {\bf 118}, 092002 (2017).
\bibitem{PRD96-032004} M.~Ablikim {\it et al.}  (BESIII Collaboration), Phys.\ Rev.\ D {\bf 96}, 032004 (2017).
\bibitem{pdg} C.~Patrignani {\it et al}.~(Particle Data Group), Chin. Phys. C {\bf 40}, 100001 (2016).
\bibitem{PRD80-091101} G.~Pakhlova {\it et al}.~(Belle Collaboration), Phys. Rev. D {\bf 80}, 091101(R) (2009).
\bibitem{supplement} See supplemental material at [URL will be inserted by publisher] for a summary of
the number of signal events, integrated luminosity, and Born cross section at each energy point.
\bibitem{bibbes3} M. Ablikim {\it et al}.~(BESIII Collaboration), Nucl. Instrum. Methods Phys. Res., Sect. A {\bf 614}, 345 (2010).
\bibitem{bibgeant4} S. Agostinelli {\it et al}.~(GEANT4 Collaboration), Nucl. Instrum. Methods Phys. Res., Sect. A {\bf 506}, 250 (2003).
\bibitem{kkmc} S.~Jadach, B.~F.~L.~Ward and Z.~Was, Comp. Phys. Commun. {\bf 130}, 260 (2000); Phys. Rev. D {\bf 63}, 113009 (2001).
\bibitem{evtgen} R.~G.~Ping, Chin. Phys. C {\bf 32}, 599 (2008); D.~J.~Lange, Nucl. Instrum. Methods Phys. Res., Sect. A {\bf 462}, 152 (2001).
\bibitem{PRD62-034003} J.~C.~Chen, G.~S.~Huang, X.~R.~Qi, D.~H.~Zhang, and Y.~S.~Zhu, Phys.\ Rev.\ D {\bf 62}, 034003 (2000).
\bibitem{PRD92-092006} M.~Ablikim {\it et al.}  (BESIII Collaboration), Phys.\ Rev.\ D {\bf 92}, 092006 (2015).
\bibitem{CPC40-063001} M.~Ablikim {\it et al.} (BESIII Collaboration), Chin. Phys. C {\bf 40}, 063001 (2016).
\bibitem{EAVS} E.~A.~Kuraev and V.~S.~Fadin,~Yad.~Fiz.~{\bf41}, 733 (1985) [Sov.~J.~Nucl.~Phys. {\bf 41}, 466 (1985)].
Since the efficiency and the ISR correction depend on the line shape of cross sections,
the measured cross sections are used as input and we iterate the procedure until the value of $(1+\delta)\epsilon$ becomes stable.
The largest variation in $(1+\delta)\epsilon$ during the last two iterations, 3.0\%, is taken as the systematic uncertainty.
\bibitem{EPJC-66-585} S.~Actis {\it et al}., Eur. Phys. J. C {\bf 66}, 585 (2010).
\bibitem{PRD89-112006} M.~Ablikim {\it et al.}  (BESIII Collaboration), Phys.\ Rev.\ D {\bf 89}, 112006 (2014).
\bibitem{PRL112-022001} M.~Ablikim {\it et al.} (BESIII Collaboration), Phys.\ Rev.\ Lett.\ {\bf 112}, 022001 (2014).
\bibitem{CPC39-093001} M.~Ablikim {\it et al.} (BESIII Collaboration), Chin. Phys. C {\bf 39}, 093001 (2015).
\bibitem{ZhuK} K.~Zhu, X.~H.~Mo, C.~Z.~Yuan, and P.~Wang, Int.\ J.\ Mod.\ Phys.\ A {\bf 26}, 4511 (2011); A.~D.~Bukin, arXiv:0710.5627. 
\bibitem{PRL100-062001} G.~Pakhlova {\it et al}. (Belle Collaboration), Phys. Rev. Lett. {\bf 100}, 062001 (2008).
\end{thebibliography}
\end{document}